\documentclass[12pt,a4paper]{article}
\usepackage{latexsym,graphicx,amssymb,amsmath}
\usepackage{setspace,bm}

\newcommand{\be}{\begin{equation}} 
\newcommand{\ee}{\end{equation}} 
\newcommand{\bea}{\begin{eqnarray}} 
\newcommand{\eea}{\end{eqnarray}} 
 
\newcommand{\rt}[1]{{}}

\makeatletter
\long\def\unmarkedfootnote#1{{\long\def\@makefntext##1{##1}\footnotetext{#1}}}
\makeatother

\begin{document}

\title{Renormalisability of the 2PI-Hartree approximation of
multicomponent scalar models in the broken symmetry phase}

\author{G. Fej\H{o}s$^{1a}$, A. Patk{\'o}s$^{2a,b}$ \vspace*{0.05cm}\\
{\it $^a$Department of Atomic Physics, E{\"o}tv{\"o}s University}
\vspace*{0.05cm}
\\
{\it $^b$Research Group for Statistical and Biological Physics}\\
{\it of the Hungarian Academy of Sciences}\\
{\it H-1117 Budapest, Hungary}
\vspace*{0.15cm}\\
Zs. Sz{\'e}p$^{3}$\vspace*{0.05cm}\\
{\it Research Institute for Solid State Physics and Optics of the}\\
{\it Hungarian Academy of Sciences, H-1525 Budapest, Hungary}
}
\date{}

\unmarkedfootnote{E-mail: $^1$geg@ludens.elte.hu,
$^2$patkos@ludens.elte.hu, $^3$szepzs@achilles.elte.hu}

\maketitle

\begin{abstract} 
Non-perturbative renormalisation of a general class of scalar field
theories is performed at the Hartree level truncation
of the 2PI effective action in the broken symmetry regime. 
Renormalised equations are explicitly constructed for the one- and
two-point functions. The non-perturbative
counterterms are deduced from the conditions for the cancellation of the
overall and the subdivergences in the complete
Hartree-Dyson-Schwinger equations, with a transparent method.
The procedure proposed in the present paper is shown to be equivalent to the iterative renormalisation method of Blaizot {\it et al.} \cite{blaizot04}.
\end{abstract}

\section{Motivation} 

One of the most popular approximation techniques in many-body quantum
theory is the Hartree approximation. In quantum field theory it
corresponds to the momentum independent two-loop truncation of the
two-particle irreducible effective action. It is used extensively both
in equilibrium \cite{cjt74,petropoulos99,lenaghan00} and
out-of-equilibrium \cite{boyanovsky95,cooper97,destri00,baacke02}
non-perturbative investigations of phase transition phenomena. Its
non-perturbative renormalisability was demonstrated as particular case
of the general proof of renormalisability of the physical quantities
computed in various 2PI approximations
\cite{heesII02,blaizot04,berges05}. These proofs are rather involved
especially in the broken symmetry phase. For this reason in many
practical applications the renormalised equations are not constructed
explicitly. For instance, investigations of the finite temperature
phase transitions in strongly interacting matter frequently either
omit zero temperature quantum corrections in the 2PI approximate
equations of the relevant 1- and 2-point functions
\cite{petropoulos99,lenaghan00,roeder03} or take into account vacuum
fluctuations by applying some cut-off \cite{roeder06}.

The exact generating 2PI-functional $\Gamma[\Phi,G]$ fulfils
generalised Ward-Takahashi identities reflecting global internal
symmetries of the models. As a consequence the 1PI effective potential
$\Gamma[\Phi,G(\Phi)]$ which arises after substituting the solution
of the stationarity condition $\delta\Gamma[\Phi,G]/\delta G=0$ at
fixed $\Phi$ generates in the broken symmetry phase an inverse
propagator vanishing for $p\rightarrow 0$. The same steps lead for a
truncated approximate $\Gamma_{tr}[\Phi,G]$ to
$\delta^2\Gamma_{tr}/\delta\Phi\delta\Phi$ which has vanishing Fourier
transform for $p\rightarrow 0$. Since this quantity in general does
not coincide with $G^{-1}(p)$ determined self-consistently,
Goldstone's theorem gets violated in approximate 2PI computations
\cite{baym77}. This failure might be one of the reasons why
2PI-Hartree approximation does not describe properly the late time
dynamics of symmetry breaking, i.e. this approximation does not lead
to a thermalised symmetry breaking ground state \cite{salle01}.
Recently a ``symmetrized'' modification was proposed to replace the
original 2PI-functional, which can be uniquely constructed from the
requirement of obeying Goldstone's theorem \cite{ivanov05a}.

A further problem concerns the renormalisability of the equations of
motion derived from the 2PI-Hartree approximation using a single
renormalised quartic coupling \cite{amelino97,lenaghan00b,destri05}.
It could be implemented consistently only in large $N$ approximations.
Renormalisability in this sense of the above referred ``symmetrized''
approximation was verified in mass-independent schemes
\cite{ivanov05b}. In parallel studies one succeeded also to implement
one-loop renormalisation-group invariance in the Hartree-like
equations \cite{destri05,destri06}.

Despite all its problems Hartree approximation remains a valuable
method in phenomenological studies when looking for the
thermodynamical behaviour of complicated multicomponent scalar models.
Our aim in the present paper is to clarify fully its renormalisability
by providing a simple and transparent explicit construction of the
counterterms for a rather general class of scalar field theories along
the method proposed in \cite{berges05,arrizabalaga06}.

The paper is organised as follows. In section~2 we present a
simplified one-step renormalisation procedure using the familiar
example of the $O(N)$ model and show its equivalence with the method
of iterative renormalisation \cite{blaizot04}. Its correct (exact)
large $N$ behaviour will be recovered. We will shortly comment on the
counter term structure of the ``symmetrized'' approximate 2PI-Hartree
functional of Ref.\cite{ivanov05a} in the present scheme. In section~3 we
introduce the general class of the models we shall investigate and
write the truncated expression of their effective quantum action
corresponding to the Hartree-approximation. In section~4 we derive the
full set of renormalisation conditions in the broken symmetry phase.
The main result is that in the 2PI-Hartree approximation there is an
interplay between the symmetry breaking pattern (an infrared feature)
and the actual set of counterterms necessary for avoiding ultraviolet
divergencies. This circumstance explains the problems signalled
earlier concerning the renormalisability of the approximation. We work
out details of the procedure for two classes of particular physical
interest. In section~5 counterterms are constructed for models which
can be embedded into the $O(N)\times O(M)$ symmetry. Section~6 is
devoted to the discussion of the $SU(N)\times SU(N)$ symmetric linear
sigma model. Throughout we pay special attention to the limit
$N\rightarrow\infty$. It will be shown for both classes of models that
in this limit the ``interference'' of the symmetry breaking pattern
with the counterterms is suppressed. Summary of our results is given
in the concluding section~7.

\section{One step renormalisation of the 2PI-Hartree approximation
 of the O(N) model in the broken phase \label{sec:one-step-renorm}} 
 
The essence of our approach can be illustrated on the example of an
$O(N)$ symmetric scalar field theory with quartic self-interaction. A
compact analysis of the model at the Hartree level of truncation
appeared in the literature already as part of a more complete
discussion of the 2PI renormalisability in Ref.~\cite{berges05}. The
treatment was shown renormalisable by constructing the necessary
counterterms iteratively. Now we show an equivalent one-step
procedure.

The 2PI-Hartree effective potential for this model is of the form \cite{cjt74}
\bea
\nonumber
V[v,G]=\frac{1}{2}\mu^2 v^2+\frac{1}{24 N} F_{abcd} v_a v_b v_c v_d
-\frac{i}{2} \int_k \ln G^{-1}_{aa}(k) 
-\frac{i}{2} \int_k \left[D^{-1}_{ab}(k) G_{ba}(k)-N\right]\\
+\frac{1}{8N} F_{abcd}\int_k G_{ab}(k)\int_p G_{cd}(p)
+V^\textnormal{ct}[v,G],
\label{Eq:ON_effpot}
\eea
where $v^2=v_a v_a,$ with $a=1\dots N$ and 
$F_{abcd}=\frac{\lambda}{3}\left(\delta_{ab}\delta_{cd}+\delta_{ac}\delta_{bd}+
\delta_{ad}\delta_{bc}\right)$ is the renormalised coupling tensor. 
The three terms in the coupling tensor 
$F_{abcd}$ are the three rank four invariants of the $O(N)$ group. 
Because in concrete calculations only the sum of the last two appears, 
it suffices to introduce only the following invariants: 
\be
t^{1}_{abcd}=\delta_{ab}\delta_{cd},\qquad
t^{2}_{abcd}=\delta_{ac}\delta_{bd}+\delta_{ad}\delta_{bc}.
\label{Eq:ON_invariants}
\ee
These invariant tensors produce two different contractions of the
indices in the fifth term of (\ref{Eq:ON_effpot}): 
$G_{aa}(k) G_{cc}(p)$ and $G_{ab}(k) G_{ab}(p).$

In the broken symmetry phase it is useful to introduce the orthogonal
projectors \cite{nemoto00}
\be
P^\sigma_{ab}=\frac{v_a v_b}{v^2},\qquad
P_{ab}^\pi=\delta_{ab}-P_{ab}^\sigma,
\label{Eq:ON_projectors}
\ee
projecting on the one dimensional ``sigma'' and $N-1$ dimensional
``Goldstone'' subspaces, respectively. (We call the degenerate sector
``Goldstone'' though in the 2PI-Hartree approximation the Goldstone
theorem is not obeyed.)
With their help the tree-level propagator in (\ref{Eq:ON_effpot}) reads as
\be
iD_{ab}^{-1}(k)=\left(k^2-\mu^2-\frac{\lambda}{6N}v^2\right)
(P_{ab}^\sigma+P_{ab}^\pi)-\frac{\lambda}{3N}v^2 P_{ab}^\sigma.
\label{Eq:ON_tree_prop}
\ee
The two terms on the right hand side of the equation correspond
to the two invariants given in (\ref{Eq:ON_invariants}).
Patterned after the tree-level propagator one writes the full
propagator in the form
\be
G_{ab}(k)=G_\sigma(k) P_{ab}^\sigma+G_\pi(k) P_{ab}^\pi,
\label{Eq:ON_full_G}
\ee
where the coefficient functions are parametrised as 
$iG_{\sigma/\pi}^{-1}(k)=k^2-M^2_{\sigma/\pi}$.

The truncation of the 2PI approximation of the quantum effective
action allows to introduce different definitions of the 4-point
function and to each one of them an independent coupling counterterm
will be associated. In Ref.~\cite{berges05} two sets of counterterms,
indexed with $A$ and $B$, were introduced corresponding to the two
independent $O(N)$ invariant structures, $t^1$ and $t^2$, building up
the coupling tensor $F_{abcd}$.

Using (\ref{Eq:ON_invariants}), (\ref{Eq:ON_tree_prop}) and
(\ref{Eq:ON_full_G}) it is an easy exercise to write the terms in the
effective potential (\ref{Eq:ON_effpot}) in terms of the exact
$\sigma$ and $\pi$ propagators keeping track of the terms which come
from contracting with the tensors $t_1$ and $t_2$. Then the following
counterterm functional can be introduced:
\be
\begin{split}
V^\textnormal{ct}[v,G_\sigma,G_\pi]=&V_4^\textnormal{ct}[v]+
V_2^\textnormal{ct}[v,G_\sigma,G_\pi]+V_0^\textnormal{ct}[G_\sigma,G_\pi],
\\
V_4^\textnormal{ct}[v]=&\frac{1}{2}\delta m_0^2
v^2+\frac{\delta\lambda_4}{24N} v^4,\\
V_2^\textnormal{ct}[v,G_\sigma,G_\pi]=&
\frac{1}{2}\left(\delta m_2^2+\frac{\delta\lambda_2^A}{6N}v^2
\right)\int_p \left(G_\sigma(p)+(N-1)G_\pi(p)\right)+
\frac{\delta\lambda_2^B v^2}{6N}\int_p G_\sigma(p),\\
V_0^\textnormal{ct}[G_\sigma,G_\pi]=&
\frac{\delta\lambda_0^A+2\delta\lambda_0^B}{24N}
\left(\int_p G_\sigma(p)\right)^2+
\frac{N-1}{24 N}\left((N-1)\delta\lambda_0^A+2\delta\lambda_0^B\right)
\left(\int_p G_\pi(p)\right)^2\\
&+\frac{\delta\lambda_0^A}{12 N}(N-1) \int_p G_\sigma(p)\int_k G_\pi(k).
\end{split}
\ee
$V_4^\textnormal{ct}$ corresponds to the classical potential and the
indices of the different counterterm functionals refer to the highest
power of the background $v$ occurring in their expression. Each $O(N)$
invariant piece of the 2PI-Hartree effective potential receives an
independent counterterm.

The stationarity conditions
\be
\frac{\delta V[v,G_\sigma,G_\pi]}{\delta G_{\sigma}(p)}=0, \qquad
\frac{\delta V[v,G_\sigma,G_\pi]}{\delta G_{\pi}(p)}=0, \qquad
\frac{\delta  V[v,G_\sigma,G_\pi]}{\delta v}=0,
\ee
give two gap equations for the pole masses and an equation of state,
which determines the vacuum condensate $v$:
\bea
M_\sigma^2&=&m_\sigma^2+\delta m_\sigma^2+
\frac{1}{6N}
\left(3\lambda+\delta\lambda_0^A+2\delta\lambda_0^B\right)T(M_\sigma^2)+
\frac{N-1}{6N}\left(\lambda+\delta\lambda_0^A\right)T(M_\pi^2),
\label{Eq:ON_sigma_gap}\\
M_\pi^2&=&m_\pi^2+\delta m_\pi^2+
\frac{1}{6N}\left(\lambda+\delta\lambda_0^A\right)T(M_\sigma^2)
+\frac{1}{6N}\left((N+1)\lambda+(N-1)\delta\lambda_0^A+
2\delta\lambda_0^B\right) T(M_\pi^2),
\label{Eq:ON_pi_gap}\\
0&=&v\left[m_\sigma^2+\delta m_0^2+\frac{\delta\lambda_4-2\lambda}{6N}v^2+
\left(3\lambda+\delta\lambda_2^A+2\delta\lambda_2^B\right)
\frac{T(M_\sigma^2)}{6N}
+\frac{N-1}{6N}\left(\lambda+\delta\lambda_2^A\right)T(M_\pi^2)
\right]\!.
\label{Eq:ON_EoS}
\eea
For compactness we have introduced the following background
dependent masses and counterterms:
\be
\begin{split}
m_\sigma^2=\mu^2+\frac{\lambda}{2 N} v^2,&\quad
\delta m_\sigma^2=\delta m_2^2 + 
\frac{\delta \lambda_2^A + 2\delta\lambda_2 ^B}{6N} v^2,\\
m_\pi^2=\mu^2+\frac{\lambda}{6 N} v^2,&\quad
\delta m_\pi^2=\delta m_2^2+\frac{\delta\lambda_2^A}{6N}v^2, 
\end{split}
\ee
and $T(M^2)=\int_p i/(p^2-M^2)$ is the tadpole integral.

Before proceeding further, we can reduce the number of counterterms.
Making use of the gap equation (\ref{Eq:ON_sigma_gap}) for the sigma
field in the equation of state (\ref{Eq:ON_EoS}), one immediately sees
that for the consistent renormalisation of the two equations one has
to require
\be
\delta m_0^2=\delta m_2^2\equiv\delta m^2\qquad
\delta\lambda_0^A=\delta\lambda_2^A\equiv\delta\lambda^A,
\qquad \delta\lambda_0^B=\delta\lambda_2^B\equiv\delta\lambda^B, \qquad
\delta\lambda_4=\delta\lambda_2^A+2\delta\lambda_2^B.
\ee
As a consequence, we are left with a very simple renormalised equation
of state:
\be
v\left[M^2_\sigma-\frac{\lambda}{3N} v^2\right]=0.
\label{Eq:ON_ren_EoS}
\ee

The key point of the renormalisation procedure applied to the two gap
equations is to know explicitly the divergence structure of the radiative
corrections, which for the tadpole diagram with cut-off regularisation
is the following:
\be
T(M^2)=\int_k\frac{i}{k^2-M^2}=\Lambda^2+T_d M^2+T_F(M^2).
\label{Eq:tadpole}
\ee
Here we used as a 4d cut-off 
$\Lambda_{\textnormal{\scriptsize CO}}=4\pi \Lambda,$
in terms of which the logarithmic divergence is 
$T_d=-\ln(e\Lambda^2_{\textnormal{\scriptsize CO}}/M_0^2)/(16\pi^2).$ 
$T_F(M^2)$ is the finite part of the tadpole integral, 
which depends also on the normalisation scale $M_0^2$.  

When substituting (\ref{Eq:tadpole}) into the two gap
equations one readily separates their finite parts: 
\bea
\label{Eq:ON_ren_sigma_gap}
M^2_\sigma&=&m^2_\sigma+\frac{\lambda}{2N} T_F(M^2_\sigma)
+(N-1)\frac{\lambda}{6N}T_F(M_\pi^2),\\
M^2_\pi&=&m^2_\pi+\frac{\lambda}{6N} T_F(M^2_\sigma)
+(N+1)\frac{\lambda}{6N}T_F(M^2_\pi).
\label{Eq:ON_ren_pi_gap}
\eea
The infinities should consistently cancel by the appropriate choice 
of the counterterms:
\bea
\label{Eq:ON_ct_sigma}
0&=&\delta m^2_\sigma+\frac{1}{6N}
\left(3\lambda+\delta\lambda^A+2\delta\lambda^B\right)
\left(\Lambda^2+T_d M^2_\sigma\right)
+\frac{N-1}{6N}\left(\lambda+\delta\lambda^A\right)
\left(\Lambda^2+T_d M^2_\pi\right)\nonumber\\
&&
+(\delta\lambda^A+2\delta\lambda^B) \frac{T_F(M^2_\sigma)}{6N}
+(N-1)\delta\lambda^A \frac{T_F(M^2_\pi)}{6N},\\
0&=&\delta m^2_\pi+\frac{1}{6N}
\left(\lambda+\delta\lambda^A\right)
\left(\Lambda^2+T_d M^2_\sigma\right)
+\frac{1}{6N}\left(\lambda(N+1)+
(N-1)\delta\lambda^A+2\delta\lambda^B\right)
\left(\Lambda^2+T_d M^2_\pi\right)\nonumber\\
&&
+\delta\lambda^A \frac{T_F(M^2_\sigma)}{6N}
+\left((N-1)\delta\lambda^A+2\delta\lambda^B\right) 
\frac{T_F(M^2_\pi)}{6N}.
\label{Eq:ON_ct_pi}
\eea

The central step of the proposed procedure consists of 
making use of the renormalised equations (\ref{Eq:ON_ren_sigma_gap})
and (\ref{Eq:ON_ren_pi_gap}) for $M_\pi^2$ and
$M_\sigma^2$ appearing in the coefficients of $T_d$. Then one can separate the
conditions for the vanishing of the overall divergence and of the
subdivergences. The former conditions do not contain any dependence on
the finite tadpole $T_F$ and read as
\be
\begin{split}
0=&\ \delta m^2_\sigma+
\frac{\Lambda^2}{6N}\left(
(N+2)\lambda+N\delta\lambda^A+2\delta\lambda^B
\right)\\&
+\frac{T_d}{6N}\left[
\left(3\lambda+\delta\lambda^A+2\delta\lambda^B\right)
m^2_\sigma
+(N-1)\left(\lambda+\delta\lambda^A\right)m^2_\pi\right],
\\
0=&\ \delta m^2_\pi+
\frac{\Lambda^2}{6N}\left(
(N+2)\lambda+N\delta\lambda^A+2\delta\lambda^B
\right)\\&
+\frac{T_d}{6N}
\left[\left(\lambda+\delta\lambda^A\right)m^2_\sigma
+\left(\lambda(N+1)+
(N-1)\delta\lambda^A+2\delta\lambda^B\right)
m^2_\pi\right].
\end{split}
\ee
The conditions for the subdivergence cancellation, which are
independent of the presence of any background, are given by the separate
vanishing of the coefficients of $T_F(M^2_\sigma)$ and
$T_F(M^2_\pi)$. It is easy to see that the two conditions 
as obtained from (\ref{Eq:ON_ct_sigma}) and (\ref{Eq:ON_ct_pi})
coincide, so one remains with
\be
\delta\lambda^{A}=-\frac{\lambda T_d}{6N}[(N+4)\lambda+(N+2)\delta 
\lambda^{A}+2\delta\lambda^{B}],\qquad
\delta\lambda^{B}=-\frac{\lambda T_d}{3N}(\lambda+ \delta\lambda^{B}). 
\label{Eq:LALB}
\ee

The cancellation of the overall divergency can be split in presence of
a background into a background dependent and a background
independent piece. The vanishing of the background independent piece
gives
\be
0=\delta m^2+\frac{1}{6N}
((N+2)\lambda+N\delta\lambda^A+2\delta\lambda^B)
\left(\Lambda^2+\mu^2 T_d\right).
\label{Eq:ON_delta_m_final}
\ee
The conditions for vanishing of the background dependent pieces are
the same as those in (\ref{Eq:LALB}). Had we not used from the
beginning the equation of state then, by comparing these conditions to
those in (\ref{Eq:LALB}) we would have obtained at this point
$\delta\lambda_2^A=\delta\lambda_0^A$ and
$\delta\lambda_2^B=\delta\lambda_0^B$.

Returning to the relation between the coupling counterterms given in 
(\ref{Eq:LALB}), one can solve the second one for $\delta\lambda^B$:
\be
\delta\lambda^B=-\frac{\lambda^2 T_d}{3N}\frac{1} {1+\frac{\lambda T_d}{3N}}.
\ee
The equation for $\delta\lambda^A$ is a bit complicated,  but 
introducing the bare coupling constants as
$\lambda_B^A=\lambda^A+\delta\lambda^A, \lambda_B^B=\lambda^B
+\delta\lambda^B$ one can rewrite the relations of (\ref{Eq:LALB}) 
in the form
\be
\begin{split}
&\frac{1}{\lambda}-\frac{1}{\lambda_B^B}=-\frac{T_d}{3N}, \\ 
&\frac{1}{\lambda}-\frac{1}{\lambda_B^A}=-\frac{T_d}{6N} 
\left[N+4+(N+2)\frac{\lambda T_d}{3N}\right]. 
\end{split}
\ee

The method of iterative renormalisation searches for the coupling
counterterms in form of infinite series.  (see {\it e.g.}
\cite{blaizot04,patkos06} for details).  When applied to the coupled
gap equations (\ref{Eq:ON_sigma_gap}) and (\ref{Eq:ON_pi_gap}) this
method leads to the following recursions for the terms of the
three counterterm series 
$\displaystyle \delta m^2=\sum_{n=1}^\infty \delta
m^{2(n)},$ $\displaystyle \delta\lambda^A=\sum_{n=1}^\infty
\delta\lambda^{A(n)}$ and $\displaystyle
\delta\lambda^B=\sum_{n=1}^\infty \delta\lambda^{B(n)}$:
\be
\begin{split} 
\delta\lambda^{A(n)}&=-\frac{\lambda T_d}{6N}[(N+2)\delta 
\lambda^{A(n-1)}+2\delta\lambda^{B(n-1)}],\\ 
\delta\lambda^{B(n)}&=-\frac{\lambda T_d}{3N}\delta\lambda^{B(n-1)}, 
\\ 
\delta m^{2(n)}&=-\frac{1}{6N}(N\delta\lambda^{A(n-1)}+2\delta 
\lambda^{B(n-1)})\left(\Lambda^2+\mu^2T_d\right),
\end{split}
\label{counter-term-iteration-ON} 
\ee
with the initial values $\delta\lambda^{A(0)}=\delta\lambda^{B(0)}=\lambda$. 
Summing up the coupling counterterms one obtains the relations 
(\ref{Eq:LALB}) and (\ref{Eq:ON_delta_m_final}) demonstrating the
equivalence of our single step approach with the iterative
construction of the counterterms.

At leading order of the large $N$ expansion all the nonvanishing coupling
counterterms are equal, since then
\begin{equation} 
\delta\lambda^A=-\frac{\lambda^2 T_d}{6} 
\frac{1}{1+\frac{\lambda T_d}{6}},\qquad \delta\lambda^B\sim{\cal O} 
\left(\frac{1}{N}\right). 
\label{Eq:ON_largeN_scaling}
\end{equation}
Note, that only in this limit one finds $\delta\lambda_4=\delta\lambda^A$.
Otherwise $\delta\lambda_4$ differs from $\delta\lambda^A$
which is a peculiar feature of the Hartree approximation.

The result (\ref{Eq:ON_largeN_scaling}) agrees with the scaling
obtained in an exact leading order large $N$ treatment of the $O(N)$
model \cite{dolan74}. This is remarkable, since an entire set of
diagrams is missing from the 2PI effective action truncated at Hartree
level (see {\it e.g.} \cite{aarts02}), which gives the one-loop bubble series
in the self consistent equation of the sigma propagator.

The same procedure can be repeated also for the ``symmetrized'' 
2PI-Hartree functional where an additional piece $\Delta V[G_{ab}]$ is added to
(\ref{Eq:ON_effpot}) enforcing the validity of Goldstone's 
theorem\cite{ivanov05a}:
\be
\Delta V[G_{ab}]=\frac{\lambda}{24N}\left[3\int_kG_{aa}(k)\int_pG_{bb}(p)-
2(N-1)\int_kG_{ab}(k)\int_pG_{ab}(p)\right].
\ee
This means that  different renormalised couplings are introduced for the
two $O(N)$ invariant 4-rank tensors, $t^\alpha_{abcd}G_{ab}G_{cd}, 
\alpha=1,2$.

The corrected gap equations for $M_\sigma^2$ and $M_\pi^2$ are readily
derived (no change is induced in the equation of state of $v$). Comparing 
the equation of state and the gap equation of the pions one finds 
that Goldstone's theorem is fulfilled under the following conditions
among the counterterms:
\begin{equation}
\delta m_0^2=\delta m_2^2,~~ 
\delta\lambda_4=\delta\lambda_2^A,
~~ \delta\lambda_0^A=\delta\lambda_2^A+2\delta\lambda_2^B,
~~ \delta\lambda_2^A=\delta\lambda_0^A+\frac{2}{N-1}\delta\lambda_0^B.
\end{equation}

In the present scheme the renormalisation of the two gap equations
results in a counterterm structure which apparently does not differ 
in any essential point from the one determined for the conventional 
Hartree truncation of the 2PI-functional, e.g. one finds different 
cut-off dependences for $\delta\lambda_0^a$ and $\delta\lambda_0^B$.
In view of the fact that for this modified functional a symmetric and
mass-independent renormalisation involving a single quartic
counterterm was proven\cite{ivanov05b}, finding the relation of the two
renormalisation schemes is an interesting task for future investigation.

The renormalisation procedure presented here is generalised to a wide class
of scalar models in the next section.

\section{2PI-Hartree approximation for multicomponent scalar \\models}

The method presented in section~2 will be applied with one notable
alteration to a general class of multicomponent scalar models
possessing various internal symmetries. The difference is that one
does not start by writing the propagators in terms of mass eigenstates
but one leaves them in matrix form. Substitution of the mass matrix
into the mass dependent divergent piece of the gap equations provides
in a single step a set of very general relations renormalisation
between various coupling and counter coupling tensors. The projection
of these equations on the different diagonal eigenblocks of the mass
matrix determines the counter couplings appearing in the counterterm
tensors as coefficients of the independent invariants. Depending on
the number of independent mass eigenvalues in the spectra it might
happen that only some combinations of them will get determined.

Let us consider the following general Lagrangian density:
\bea
L=&&\frac{1}{2}[\partial_\mu\sigma_a\partial^\mu\sigma_a+
\partial_\mu\pi_\alpha\partial^\mu\pi_\alpha-\mu_S^2\sigma_a\sigma_a
-\mu_P^2\pi_\alpha\pi_\alpha]\nonumber\\
&&-\frac{1}{3}F_{abcd}^S\sigma_a\sigma_b\sigma_c\sigma_d-\frac{1}{3}
F_{\alpha\beta\gamma\delta}^P\pi_\alpha\pi_\beta\pi_\gamma\pi_\delta
-2H_{\alpha\beta,cd}\pi_\alpha\pi_\beta\sigma_c\sigma_d.
\label{Eq:general_Lagrangian}
\eea
This model includes variants of $O(N)$ symmetric models and also 
the  $SU(N)_L\times SU(N)_R$ symmetric matrix model
when specific expressions are chosen for the coefficient tensors
$F_{\alpha\beta\gamma\delta}^P, F_{abcd}^S$ and $H_{\alpha\beta,cd}$, 
which reflect the structure of the group
algebra. 
Assuming that in the broken symmetry phase only the $\sigma_a$ fields 
acquire an expectation value $v_a$ one obtains the following
tree-level inverse propagators:
\be
iD_{S,ab}^{-1}(k)=(k^2-\mu^2_S)\delta_{ab}-4F^S_{abcd}v_c v_d,\qquad
iD_{P,\alpha\beta}^{-1}(k)=(k^2-\mu^2_P)\delta_{\alpha\beta}
-4H_{\alpha\beta,cd}v_c v_d.
\ee

We introduce the economical compact ``hypervector'' notations:
\be
\begin{split}
&\delta_{AB}=\begin{pmatrix}
\delta_{ab}\\ \delta_{\alpha\beta}
\end{pmatrix},\ 
\mu^2_{AB}=\begin{pmatrix}
\mu_S^2 \delta_{ab}\\ \mu_P^2 \delta_{\alpha\beta}
\end{pmatrix},\ 
D_{AB}^{-1}=
\begin{pmatrix} D_{S,ab}^{-1} \\ D_{P,\alpha \beta}^{-1}
\end{pmatrix},\ 
G_{AB}=\begin{pmatrix} G_{S,ab}\\ G_{P,\alpha\beta}
\end{pmatrix},\\
&\bar\sigma_A=
\begin{pmatrix}
v_a\\0
\end{pmatrix},\ 
\bar\sigma_A\bar\sigma_B=
\begin{pmatrix}
v_a v_b\\0
\end{pmatrix},
\end{split}
\ee
where $G_{P,S}$ are the exact propagator matrices of the 2PI approximation.
We also organise the coefficient tensors of
(\ref{Eq:general_Lagrangian}) in the following hypermatrix: 
\be
Q_{ABCD}=
\begin{pmatrix}
F_{abcd}^S\  & H_{ab,\gamma\delta}\\
H_{\alpha\beta,cd}\  & F^P_{\alpha\beta\gamma\delta}
\end{pmatrix}.
\label{Eq:hypermatrix_Q}
\ee
This hypermatrix inherits from the component tensors 
the following symmetries under index permutations: 
\be
Q_{A B C D}=Q_{C D A B}=Q_{B A C D}
\ee
and all combinations of these transformations. 

In the notation introduced above,
the 2PI-Hartree effective potential for the models defined by 
(\ref{Eq:general_Lagrangian}) reads as
\bea
\nonumber
V[\sigma_A,G_{AB}]&=&\frac{1}{2}(\mu_{AB}^2)^\textnormal{T} 
\bar\sigma_A\bar\sigma_B
+\frac{1}{3} (\bar\sigma_A\bar \sigma_B)^\textnormal{T} 
Q_{ABCD} (\bar\sigma_C\bar\sigma_D)
-\frac{i}{2} \int_k\left[(D_{AB}^{-1}(k))^\textnormal{T}G_{BA}(k)
-\delta_{AB}^\textnormal{T}\delta_{BA}\right]
\\&&
-\frac{i}{2} \textnormal{Tr}\int_k (u\otimes \ln G_{AB}^{-1}(k))
+\int_k G_{AB}^\textnormal{T}(k) Q_{ABCD} \int_p G_{CD}(p)+
V^\textnormal{ct}[\sigma_A,G_{AB}],
\eea
where usual matrix operations were used,
$u=\begin{pmatrix}1\\1\end{pmatrix}$, $\otimes$ denotes dyadic product
and the transpose ($\textnormal{T}$) is to be understood as acting on
the blocks of hypermatrices but not within the blocks.

The counterterms are introduced by the method of
\cite{berges05,arrizabalaga06} generalising the case of the $O(N)$
model presented in section~\ref{sec:one-step-renorm}. One has 
\bea
\nonumber
V^\textnormal{ct}[\sigma_A,G_{AB}]&=&
V^\textnormal{ct}_4[\sigma_A]+V^\textnormal{ct}_2[\sigma_A,G_{AB}]+
V^\textnormal{ct}_0[G_{AB}],\\
V^\textnormal{ct}_4[\sigma_A]&=&\frac{1}{2}
(\delta\tilde\mu_{AB}^2)^\textnormal{T} \bar\sigma_A\bar\sigma_B
+\frac{1}{3} (\bar\sigma_A\bar \sigma_B)^\textnormal{T} 
\delta\tilde Q_{ABCD} (\bar\sigma_C\bar\sigma_D),
\label{Eq:Gen_V4_ct}
\\
V^\textnormal{ct}_2[\sigma_A,G_{AB}]&=&\frac{1}{2}
(\delta\hat\mu_{AB}^2)^\textnormal{T}\int_k G_{BA}(k)+
4(\bar\sigma_C\bar\sigma_D)^\textnormal{T}
\delta \hat Q_{ABCD}^\textnormal{T}\int_k G_{BA}(k),
\label{Eq:Gen_V2_ct}
\\
V^\textnormal{ct}_0[G_{AB}]&=&\int_k G_{AB}^\textnormal{T}(k) 
\delta Q_{ABCD} \int_p G_{CD}(p).
\label{Eq:Gen_V0_ct}
\eea
Here, we used the freedom to introduce three different coupling counter tensors
$\delta Q$, $\delta\hat Q$ and $\delta \tilde Q$ for the three 
different definitions of the 4-point couplings
as well as two mass counter terms $\delta \hat \mu^2$ and $\delta \tilde \mu^2$
for the two different definitions of the 2-point couplings 
\cite{berges05,arrizabalaga06}. For the symmetry breaking pattern considered,
in terms of the components, the counter couplings read as
\bea
\delta \hat \mu^2_{AB}=\begin{pmatrix}
\delta\hat\mu^2_S\delta_{ab}\\\delta\hat\mu^2_P\delta_{\alpha\beta} 
\end{pmatrix},
\delta \tilde \mu^2_{AB}=\begin{pmatrix}
\delta\tilde\mu^2_S\delta_{ab}\\ 0
\end{pmatrix},
\delta Q=
\begin{pmatrix}
\delta F^S & \delta H\\
\delta H  & \delta F^P
\end{pmatrix},
\delta \hat Q=
\begin{pmatrix}
\delta \hat F  & 0\\
\delta \hat H  & 0
\end{pmatrix},
\delta \tilde Q=
\begin{pmatrix}
\delta \tilde F & 0\\
0  & 0
\end{pmatrix},
\label{Eq:countertensors_general}
\eea
with the same index structure inside each block like
in (\ref{Eq:hypermatrix_Q}). All the components of the hypermatrices
are linear combinations of rank-4 invariant tensors of the
symmetry group considered. 

The equations for the full propagator $G_{AB}$ and the vacuum
expectation value $v_a$ follow from the stationarity conditions 
$\delta V/\delta G_{AB}=0$ and $\delta V/\delta v_a=0$. 
Because of the momentum independence of the self-energy in the
Hartree approximation, one can write
\be
iG_{AB}^{-1}(k)=k^2 \delta_{AB}^\textnormal{T} -
(M_{AB}^2)^\textnormal{T},\qquad
(M_{AB}^2)^\textnormal{T}=(M_{ab}^2,M_{\alpha \beta}^2),
\ee
where $M_{ab}^2,$ and $M_{\alpha \beta}^2$ are the exact squared mass matrices
in the `S' and `P' sectors, respectively. 
With this parametrisation the selfconsistent gap
equations and the equation of state are written as
\bea
M^2_{AB}&=&m^2_{AB}+4Q_{ABCD}\int_k G_{AB}(k)+\delta \hat m^2_{AB}
+4\delta Q_{ABCD}\int_k G_{CD}(k),
\label{gap-equations}
\\
0&=&\sigma_B^T\left[
\mu^2_{AB}+\delta\tilde\mu_{AB}^2+
\frac{4}{3}(Q_{ABCD}+\delta\tilde Q_{ABCD})\bar\sigma_C\bar\sigma_D+
4(Q_{ABCD}+\delta\hat Q^\textnormal{T}_{ABCD})\int_k G_{CD}(k)\right].
\label{Eq:EoS-general}
\eea
In (\ref{gap-equations}) some convenient short-hand notations were introduced:
\be
m_{AB}^2=\mu^{2}_{AB}+4Q_{ABCD} \bar\sigma_C\bar\sigma_D, \qquad 
\delta \hat m_{AB}^2=\delta\hat\mu^{2}_{AB}+
4\delta \hat Q_{ABCD}\bar\sigma_C\bar\sigma_D.
\ee
  
\section{General renormalisation conditions for the 2PI-Hartree approximation}

The real and symmetric exact squared mass matrices in the `S' and `P'
sectors can be diagonalised with orthogonal transformations:
\be 
O_{ac}^S M^2_{ab}O_{bd}^S=\tilde M^2_c\delta_{cd},\qquad
O_{\alpha \gamma}^P M^2_{\alpha\beta}O_{\beta\delta}^P=
\tilde M^2_\gamma\delta_{\gamma\delta},
\ee
where on the right hand side there is no summation over the repeated indices.
With this transformation also the propagator matrices become diagonal
and the corresponding tadpole integrals can be evaluated explicitly.
Using hypervectors one introduces the following short-hand notations
\be
M^2_{CD}=O_{CE} O_{DE} \tilde M^2_E,\qquad  
\int_k G_{CD}(k)=O_{CE} O_{DE} T(\tilde M^2_E),
\label{Eq:OOmeaning}
\ee
with the understanding that in the upper component of the squared mass and
propagator hypervectors one diagonalises with $O^S$ and in the lower
one with $O^P$. One makes explicit the divergent piece
of the tadpole integral by writing
\be
T(\tilde M^2_E)=\Lambda^2 u + \tilde M_E^2 T_d +T_F(\tilde M^2_E).
\label{Eq:tadpole_decomp_general}
\ee


With help of (\ref{Eq:OOmeaning}) and
(\ref{Eq:tadpole_decomp_general}) one expresses the integrals of
(\ref{gap-equations}) in terms of tadpole integrals of the propagating
eigenmodes and obtains for the gap equations
\bea
\nonumber
M^2_{AB}&=&\mu^2_{AB}+\delta\hat\mu^2_{AB}
+4(Q_{ABCD}+\delta \hat Q_{ABCD})\bar\sigma_C\bar\sigma_D\\
&&+4(Q_{ABCD}+\delta Q_{ABCD})\left[
\Lambda^2\delta_{CD}+M_{CD}^2 T_d
+O_{CE} O_{DE} T_F(\tilde M_E^2)
\right].
\label{Eq:gap-equations-tad}
\eea
The renormalised gap equations are easily extracted from the above
sum by separating its finite pieces:
\be
M^2_{AB}=\mu^2_{AB}+4Q_{ABCD}\bar\sigma_C\bar\sigma_D
+4Q_{ABCD}O_{CE} O_{DE} T_F(\tilde M_E^2).
\label{ren-gap}
\ee  

One has to choose the counter tensors appropriately ensuring the vanishing of
all independent overall divergences and subdivergences in
(\ref{Eq:gap-equations-tad}). Substituting for the squared mass matrix
$M_{CD}^2$ its expression from the renormalised gap equation
(\ref{ren-gap}) divergence cancellation imposes the following relation 
on the counterterms
\bea
0=&&\delta\hat\mu^2_{AB}+4(Q_{ABCD}+\delta Q_{ABCD})
\left(\Lambda^2\delta_{CD}+\mu^2_{CD} T_d\right)
\nonumber\\
&&
+4\left[\delta\hat Q_{ABMN}+4T_d(Q_{ABCD}+\delta Q_{ABCD})Q_{CDMN}\right]
\bar\sigma_M\bar\sigma_N
\nonumber
\\
&&+4\left[\delta Q_{ABMN}+4T_d(Q_{ABCD}+\delta Q_{ABCD})Q_{CDMN}\right]
O_{ME} O_{NE} T_F(\tilde M_E^2).
\label{renorm-cond}
\eea 
Note, that we split the overall divergence into two sets (the first two
lines on the right hand side of (\ref{renorm-cond})), one independent of
the background and another depending on it quadratically.

Turning now to the renormalisation of the equation of state we first
express $\mu^2_{AB}$ from (\ref{gap-equations}) and substitute it into
(\ref{Eq:EoS-general}). Then one does the same steps as in the case of
the gap equations: rewrite the integrals in terms of tadpoles of mass
eigenstates using (\ref{Eq:OOmeaning}), separate the divergent part of
the tadpole integral using (\ref{Eq:tadpole_decomp_general}) and make
use of the finite gap equations (\ref{ren-gap}).
The finite equation of state has a very simple form 
(compare to (\ref{Eq:ON_ren_EoS})):
\be
\bar\sigma_B^\textnormal{T}\left[M_{AB}^2-\frac{8}{3}Q_{ABCD} 
\bar\sigma_C\bar\sigma_D \right]=0.
\ee
The condition for vanishing of all the overall divergences and
subdivergences in the equation of state gives 
\bea
0=&&\bar\sigma_B^\textnormal{T}\left\{
\delta\tilde \mu^2_{AB}-\delta\hat\mu^2_{AB} +
4(\delta\hat Q_{ABCD}^\textnormal{T}-\delta Q_{ABCD})
\left(\Lambda^2\delta_{CD}+\mu^2_{CD} T_d\right)\right.
\nonumber\\
&&
+4\left[\frac{1}{3}\delta\tilde Q_{ABMN}-\delta\hat Q_{ABMN}
+4T_d(\delta \hat Q_{ABCD}^\textnormal{T}-\delta Q_{ABCD})
Q_{CDMN}\right] \bar\sigma_M\bar\sigma_N
\nonumber
\\
&&\left.
+4(\delta\hat Q_{ABCD}^\textnormal{T}-\delta Q_{ABCD})
\left[I_{CM} I_{DN}+ 4 T_d Q_{CDMN}\right] 
O_{ME} O_{NE} T_F(\tilde M_E^2)\right\},
\label{Eq:renorm-cond-EoS}
\eea
where in the last line we introduced 
$I_{CM}=\textrm{diag}(\delta_{cm},\delta_{\gamma\mu})$.

The counter tensors enter in three different types of combinations in
(\ref{renorm-cond}) and (\ref{Eq:renorm-cond-EoS}). The first line in
these equations is independent on the background. The second line
reflects the presence of background dependent overall divergences. The
expressions in the third line are due to the presence of
subdivergences. These latter expressions are products of a piece
independent of the pole masses and another one which through its
$\tilde M_E^2$ dependence is potentially dependent on the temperature,
the chemical potential and other ``environmental'' parameters.
Renormalisability of the approximation is equivalent to ensure the
vanishing of all three types of expressions without imposing
environment dependent conditions.

A sufficient condition for vanishing of the subdivergences contained
in the third line of (\ref{Eq:renorm-cond-EoS}) is to choose the same
coupling counter tensor in (\ref{Eq:Gen_V2_ct}) and
(\ref{Eq:Gen_V0_ct}), that is 
\be
\delta\hat Q_{ABCD}^\textnormal{T}=\delta Q_{ABCD}.
\label{Eq:choise}
\ee
Then the first line of (\ref{Eq:renorm-cond-EoS}) gives
$\delta\tilde \mu_S^2=\delta\hat\mu_S^2$. 
The vanishing of background dependent divergences of the
second line in (\ref{Eq:renorm-cond-EoS}) requires
\be
\bar\sigma_B^\textnormal{T}
\left(\frac{1}{3}\delta\tilde Q_{ABMN}-\delta\hat Q_{ABMN}\right)
\bar\sigma_M\bar\sigma_N=0,
\label{eos-gap-compatibility}
\ee
which in view of (\ref{Eq:countertensors_general}) and
(\ref{Eq:choise}) is equivalent to
\be
\left(\frac{1}{3}\delta \tilde F_{abmn}-
\delta F_{abmn}\right)v_b v_m v_n=0.
\label{Eq:compatibility_condition}
\ee

The vanishing of the overall divergence in the gap equations requires
the separate cancellation of the first line in (\ref{renorm-cond}):
\be  
\delta \hat\mu^2_{AB}+4(Q_{ABCD}+\delta Q_{ABCD})\left(\Lambda^2
\delta_{CD}+\mu^2_{CD}T_d\right)=0.
\label{overall-cancel}
\ee

The condition for vanishing of the subdivergences demands
the vanishing of the coefficient of the finite $\tilde M_E^2$-dependent
combination of the tadpole integrals appearing in the third line of
(\ref{renorm-cond}).  An obvious sufficient condition is to impose the
vanishing of the hypermatrix in the square bracket of this line for
all its tensorial components. This would yield the following set of
linear equations for $\delta Q_{ABCD}$:
\be
\delta Q_{ABMN}+4T_d(Q_{ABCD}+\delta Q_{ABCD})Q_{CDMN}=0.
\label{4coupling-conditions}
\ee

Due to the choice in (\ref{Eq:choise}), in the subspace of the sector
`S' spanned by the nonzero components $v_m$ of the background, the
above condition is the same as the one coming from the vanishing of
the background dependent overall divergences (second line in
(\ref{renorm-cond})). Still, it is unnecessary to impose this
condition on elements for which the mass-dependent term, e.g. $O_{ME}
O_{NE}T_F(\tilde M_E^2)$ vanishes. Taking into account the
block-diagonal form of the matrix $M_{AB}^2$,
the matrix condition should be fulfilled within each coupled mass block
separately. Degenerate modes of common mass (like the Goldstone modes)
also form a sector sharing common $T_F(\tilde M^2)$. In this block the
product $O_{ME}O_{NE}$ gives a projector $P_{MN}$ onto this subspace.
The form of renormalisation condition to be applied on this subspace
arises by multiplying (\ref{4coupling-conditions}) by $P_{MN}$.

Let us analyse first the consequences of (\ref{4coupling-conditions})
in the symmetric phase. There is only one completely degenerate block
in both sectors `S' and `P'. In this case one has
\be
(M_{AB}^2)^\textnormal{T}=(M^2_S\delta_{ab},M^2_P\delta_{\alpha\beta}),
\qquad (G_{AB}(k))^\textnormal{T}=
(G_S\delta_{ab}(k),G_P\delta_{\alpha\beta}(k),
\ee
hence one pair of indices of the coupling tensors will be contracted. 

Taking for convenience the trace in the original gap equation
(\ref{gap-equations}) separately in sectors `S' and `P', one
arrives at the following renormalisation condition for the vanishing
of the subdivergences
\be
\delta Q+4T_d (Q+\delta Q)Q=0,
\label{Eq:renorm_cond_symphase}
\ee
where we have introduced $Q=Q_{aa\gamma\gamma},$ and $\delta Q=\delta
Q_{aa\gamma\gamma}$. 
This equation actually determines three scalar counterterms $\delta F^P,
\delta F^S$ and $\delta H$. Note, that these contracted tensor
couplings enter in the renormalisation condition for vanishing of the
overall divergences (\ref{overall-cancel}). As we will see in
concrete examples in the symmetric phase only a combination of the
coupling counterterms is determined. This combination is split in the
broken symmetry phase.
 
For the concrete applications of the next sections we write in a less
compact notation the conditions for vanishing of the subdivergences
coming from the last line of (\ref{renorm-cond}):
\be
\begin{split}
&
\left\{
\delta F^S_{abmn}\, +4T_d[(F^S_{abcd}+\delta F^S_{abcd})F^S_{cdmn}+
(H_{ab,\gamma\delta}+\delta H_{ab,\gamma\delta})H_{\gamma\delta,mn}]
\right\}O^S_{me}O^S_{ne} T_F(\tilde M_{S,e}^2)=0,
\\
&
\left\{
\delta H_{ab,\mu\nu}\, +4T_d[(F^S_{abcd}+\delta F^S_{abcd})H_{cd,\mu\nu} 
+(H_{ab,\gamma\delta}+\delta H_{ab,\gamma\delta})F^P_{\gamma\delta\mu\nu}]
\right\}O^P_{\mu\epsilon}O^P_{\nu\epsilon} T_F(\tilde M_{P,\epsilon}^2)=0,
\\
&
\left\{
\delta H_{\alpha\beta,mn}+4T_d[(F^P_{\alpha\beta\gamma\delta}
+\delta F^P_{\alpha\beta\gamma\delta})H_{\gamma\delta,mn}
+(H_{\alpha\beta,cd} +\delta H_{\alpha\beta,cd})F^S_{cdmn}]
\right\}O^S_{me}O^S_{ne} T_F(\tilde M_{S,e}^2)=0,
\\
&
\left\{
\delta F^P_{\alpha\beta\mu\nu}\, +4T_d[(F^P_{\alpha\beta\gamma\delta}
+\delta F^P_{\alpha\beta\gamma\delta})F^P_{\gamma\delta\mu\nu}+
(H_{\alpha\beta,cd}+\delta H_{\alpha\beta,cd})H_{cd,\mu\nu}]
\right\}O^P_{\mu\epsilon}O^P_{\nu\epsilon} T_F(\tilde M_{P,\epsilon}^2)=0.
\end{split}
\label{detailed-conditions}
\ee

One sees that products of two rank-4 tensors contracted with two pairs
of indices are involved. The coupling and counter coupling tensors are
linear combinations of independent rank-4 invariant tensors $t^\alpha$
of a given group:
\be
F^{P/S}_{abcd}=\sum_\alpha f_\alpha^{P/S} t^\alpha_{abcd}, \quad 
H_{ab,cd}=\sum_\alpha h_\alpha t^\alpha_{abcd},\quad
\delta F^{P/S}_{abcd}=\sum_\alpha \delta f_\alpha^{P/S} t^\alpha_{abcd}, \quad
\delta H_{ab,cd}=\sum_\alpha\delta h_\alpha t^\alpha_{abcd}.
\ee 
The product of the tensors appearing in (\ref{detailed-conditions})
can be conveniently worked out in form of a multiplication table for
the invariants:
\be
t^\alpha_{abcd}t^\beta_{cdef}=\sum_\gamma g_{\alpha\beta\gamma}
t^\gamma_{abef}.
\ee
After projecting the resulting equations onto a given coupled block of
the original gap equations one equates the coefficients of the arising
independent (tensorial) expressions and determines the coupling
counterterms $\delta f^{P/S}_\alpha, \delta h_\alpha$.  
These steps will be explicitly performed next for some concrete models
of physical interest.
 
\section{Analysis of the $O(N)\times O(M)$ symmetric model}

The application of the renormalisation conditions to the $O(N)\times O(M)$
symmetry structure proceeds by specifying the invariant tensor
structure (recall Eq.~(\ref{Eq:ON_invariants})) of the coupling and 
coupling counterterm tensors appearing
in (\ref{Eq:hypermatrix_Q}) and (\ref{Eq:countertensors_general}) 
(the `S' sector will be associated with the $O(N)$ symmetry):
\begin{alignat}{2}
\nonumber
\displaystyle F^S_{abcd}&=\lambda^S(t^1_{abcd}+t^2_{abcd}), &\qquad  
\displaystyle F^P_{\alpha\beta\gamma\delta}&=\lambda^P
(t^1_{\alpha\beta\gamma\delta}+t^2_{\alpha\beta\gamma\delta}),\\
\delta F^S_{abcd}&=\delta\lambda^S_A t^1_{abcd}+
\delta\lambda^S_B t^2_{abcd}, &\qquad 
\delta F^P_{\alpha\beta\gamma\delta}&=\delta\lambda^P_A
t^1_{\alpha\beta\gamma\delta}+\delta\lambda^P_B
t^2_{\alpha\beta\gamma\delta},
\label{Eq:ON_tensors_def}
\\
H_{ab,\gamma\delta}&=\lambda^H t^1_{ab\gamma\delta},&\qquad 
\delta
H_{ab,\gamma\delta}&=\delta\lambda^H t^1_{ab\gamma\delta}.
\nonumber
\end{alignat}
Written in an obvious compact notation, the rank-4 $O(N)$ invariant
tensors obey simple multiplication rule (a pair of indices is contracted):
\be
t^1*t^1=Nt^1,\qquad t^1*t^2=2t^1,\qquad t^2*t^2=2t^2.
\label{Eq:ON_multiplication_table}
\ee
Similar multiplication table holds for the $O(M)$ invariants of the
`P' sector.

There are three blocks generated by (\ref{detailed-conditions}) in
which the equations for the coupling counter tensors are to be
satisfied separately. Since the symmetry breaking occurs in the $O(N)$
sector, it is easy to check that in this sector one has $N-1$
``Goldstone'' modes with mass $\tilde M_\pi$ and one massive mode with
mass $\tilde M_\sigma$. The $O(M)$ sector remains fully degenerate, 
all modes have the common mass $\tilde M_P$.

In this latter `P'-sector one has (see (\ref{detailed-conditions}))
\be
O^P_{\mu\epsilon} O^P_{\nu\epsilon} 
T_F(\tilde M^2_{P,\epsilon})=T_F(\tilde M_P^2) 
O^P_{\mu\epsilon} O^P_{\nu\epsilon}=T_F(\tilde M_P^2)\delta_{\mu\nu}.
\ee
This means that we have to take the trace with respect two the
free Greek indices in the curly bracket of the second and fourth equation of
(\ref{detailed-conditions}). Equating with zero the coefficients of the 
resulting two tensorial structures $\delta_{ab}$ and 
$\delta_{\alpha\beta}$ one obtains:
\be
\begin{split}
(M+2)\delta\lambda^P=&-4T_d\left[(M+2)^2\lambda^P
\left(\lambda^P+\delta\lambda^P\right)+
MN\lambda^H(\lambda^H+\delta\lambda^H)\right],\\
\delta\lambda^H=&-4T_d\left[(N+2)\lambda^H\left(\lambda^S+
\delta\lambda^S\right)+(M+2)\lambda^P(\lambda^H+\delta\lambda^H)
\right],
\end{split}
\label{Eq:ON_P_sector_counterterms}
\ee
where we introduced the notations 
$(M+2)\delta\lambda^P\equiv M\delta\lambda^P_A+2\delta\lambda^P_B,
(N+2)\delta\lambda^S\equiv N\delta\lambda^S_A+2\delta\lambda^S_B$.

In the `S'-sector using (\ref{Eq:ON_tensors_def}) 
and (\ref{Eq:ON_multiplication_table}) in the first and third equation of
(\ref{detailed-conditions}) one obtains for the tensors in the curly brackets
of these equations the following expressions
\be
\begin{split}
\{...\}=&\ t^1_{abmn}\left\{\delta\lambda_A^S+4T_d\left[
\lambda^S\left((N+4)\lambda^S+(N+2)\delta\lambda_A^S+2\delta\lambda_B^S
\right)+M\lambda^H(\lambda^H+\delta\lambda^H)
\right]
\right\}\\
&+t^2_{abmn}\left[
\delta\lambda_B^S+8 T_d \lambda^S(\lambda^S+\delta\lambda_B^S)
\right],\\
\{...\}=&\ t^1_{\alpha\beta mn}\left\{
\delta\lambda^H+4T_d\left[
(M+2)\lambda^H\left(\lambda^P+
\delta\lambda^P\right)
+(N+2)\lambda^S(\lambda^H+\delta\lambda^H)
\right]
\right\}.
\end{split}
\label{Eq:ON-S_sector}
\ee
With help of the projectors given in 
(\ref{Eq:ON_projectors}) one writes
\be
O^S_{me} O^S_{nf} T_F(\tilde M^2_{S,e})\delta_{ef}=
P_{mn}^\sigma T_F(\tilde M_\sigma^2)+P_{mn}^\pi T_F(\tilde M_\pi^2).
\ee
This means that one has to project both expressions in 
(\ref{Eq:ON-S_sector}) on the $1$-dimensional and 
the $N-1$ dimensional degenerate parts of the spectra by applying 
the corresponding projectors and then equate the resulting expressions to
zero.
Applying this procedure to the first expression of (\ref{Eq:ON-S_sector})
gives the following two relations between the counterterms:
\be
\begin{split}
\delta\lambda^S_A=&-4T_d\Big[\lambda^S
\big((N+4)\lambda^S+(N+2)\delta \lambda^S_A+2\delta\lambda^S_B\big)
+M\lambda^H(\lambda^H+\delta\lambda^H)\Big],\\
\delta\lambda^S_B=&-8 T_d\lambda^S\left(\lambda^S+
\delta\lambda^S_B\right).
\end{split}
\label{on-sector}
\ee
We have used that the projection of the rank-4 tensors can
be expressed with help of $P^\sigma$ and $P^\pi$.

The equations for $\delta\lambda^S_A$ and $\delta\lambda^S_B$ if
summed with appropriately chosen coefficients reproduce the ``mirror''
of the equation found for $(M+2)\delta\lambda^P$ in the
$O(M)$-symmetric P-sector: $S\leftrightarrow P$ and $N\leftrightarrow
M$. Here, however, we have two separate tensor structures, which
determine $\delta\lambda^S_A,\delta\lambda^S_B$, and not only the
combination $\delta\lambda^S$. This feature shows that the
renormalisability of the 2PI-Hartree approximation requires different
counterterm structure in the symmetric case and when one deals with
broken symmetry, due to the spectral structure induced by the
specific symmetry breaking pattern.

Since $P_{mn}^\sigma t^1_{\alpha\beta mn}=\delta_{\alpha\beta}$ and
$P_{mn}^\pi t^1_{\alpha\beta mn}=(N-1)\delta_{\alpha\beta}$,
upon projecting the second equation of (\ref{Eq:ON-S_sector}) 
on the two blocks of the `S'-sector one finds
\be
\delta\lambda^H=-4T_d\Big[(M+2)\lambda^H\left(\lambda^P+
\delta\lambda^P\right)
+(N+2)\lambda^S(\lambda^H+\delta\lambda^H)\Big].
\label{Eq:DL_H_Sigma}
\ee
This expression is again the ``mirror''of that in the second
line of (\ref{Eq:ON_P_sector_counterterms}) under the interchange 
$S\leftrightarrow P$ and $N\leftrightarrow M$. 
Its explicit expression can be obtained either from
(\ref{Eq:DL_H_Sigma}) and the first line of 
(\ref{Eq:ON_P_sector_counterterms}) or the second line of
(\ref{Eq:ON_P_sector_counterterms}) and the sum of the two equations
in (\ref{on-sector}). The resulting $\delta\lambda^H$ is 
symmetric, as expected, under the interchange $S\leftrightarrow P$
and $N\leftrightarrow M$:
\be
\delta\lambda^H=-\frac{4T_d\lambda^H U}{1+4T_d U},\ \ \quad
U=(M+2)\lambda^P+(N+2)\lambda^S+4T_d \left[
(M+2)(N+2)\lambda^S\lambda^P-MN(\lambda^H)^2
\right].
\label{Eq:DL_H_sol}
\ee

The compatibility condition (\ref{Eq:compatibility_condition}),
obtained by confronting the gap equations and the equation of state,
imposes condition only on the `S'-sector. By the symmetry of the
classical potential one puts
\be
\delta\tilde F_{abmn}^S=\delta\tilde\lambda(\delta_{ab}
\delta_{mn}+\delta_{am}\delta_{bn}+\delta_{an}\delta_{bm}),
\ee
which leads to the relation
\be
\delta\tilde\lambda=\delta\lambda_A^S+2\delta\lambda_B^S.
\label{on-counter-coupling-rel}
\ee

In summary we see that there are four independent coupling counterterms
$\delta\lambda^S_A,$ $\delta\lambda^S_B,$ $\delta\lambda^P,$
$\delta\lambda^H$ which renormalise the
2PI-Hartree approximation of the Dyson-Schwinger equations of this
model in the broken symmetry phase. Apart from these, there are two
mass counterterms which can be easily obtained from (\ref{overall-cancel}).

\paragraph{Special cases}

One recovers the $O(N)$ model from the general Lagrangian
(\ref{Eq:general_Lagrangian}) if only one series of fields appears, that 
is in (\ref{Eq:ON_tensors_def}) one has 
\be
F^P_{\alpha\beta\gamma\delta}=\delta F^P_{\alpha\beta\gamma\delta}
=H_{ab,\gamma\delta}=\delta H_{ab,\gamma\delta}=0.
\ee
The two equations of (\ref{on-sector})
reduce to the following relations between the two counter-couplings
$\delta\lambda_A,\delta\lambda_B$ ($\lambda^S\equiv\lambda$):
\be
\begin{split}
\delta\lambda_A=&-4\lambda T_d
\left[(N+4)\lambda+(N+2)\delta\lambda_A+2\delta\lambda_B\right],\\
\delta\lambda_B=&-8\lambda T_d\left[\lambda+\delta\lambda_B\right].
\end{split}
\ee
These equations exactly reproduce those in (\ref{Eq:LALB}) 
after rescaling the couplings and the counterterms by $24 N$. Equation
(\ref{on-counter-coupling-rel}) applies in unchanged form.

Our model can accommodate also the case of two interacting $N$-plets,
if one interprets the fields $\pi_a$ as a second $N$-plet with 
$O(N)$ invariant selfcoupling and assuming that the interaction term 
between them is unchanged.
Its analysis is similar to the general $O(N)\times O(M)$ case just one
chooses $N=M$.  The coupling counterterms relevant in the large $N$
limit are obtained from the first equations of
(\ref{Eq:ON_P_sector_counterterms}) and (\ref{on-sector}), and
from (\ref{Eq:DL_H_Sigma}). Using the notation
$\lambda^S_A\equiv\lambda^S, \lambda^P_A\equiv\lambda^P$ one has:
\be
\begin{split}
\delta\lambda^P=&-4T_dN\left[\lambda^P\left(
\lambda^P+\delta\lambda^P\right)+\lambda^H(\lambda^H+\delta
\lambda^H)\right],\\
\delta\lambda^S=&-4T_dN\left[\lambda^S\left(
\lambda^S+\delta\lambda^S\right)+\lambda^H(\lambda^H+\delta
\lambda^H)\right],\\
\delta\lambda^H=&-4T_dN\left[\lambda^H\left(\lambda^S
+\delta\lambda^S
\right)+\lambda^P(\lambda^H+\delta\lambda^H)\right].
\end{split}
\ee
The solution for $\delta \lambda^H$  is obtained from (\ref{Eq:DL_H_sol})
by taking $M=N$ and making the replacement $N+2\to N$. 
In terms of $\delta \lambda^H$ the solution for $\delta
\lambda^P$ and $\delta \lambda^S$ can be readily obtained.

\section{The $SU(N)\times SU(N)$ meson model in 2PI-Hartree 
approximation}
For $N=3$ this model is one of the most popular effective meson
models. To our knowledge, the proper renormalisation of its
2PI-Hartree approximation appears here for the first time
in the literature.

The four-point coupling tensors describing the
self-interaction of scalars and pseudoscalar mesons and the
interaction between them is usually written in the following form
\cite{haymaker73,lenaghan00}:
\be
\begin{split}
F_{abcd}^S=F_{abcd}^P=&\ 
\frac{g_1}{4}(\delta_{ab}\delta_{cd}+\delta_{ac}\delta_{bd}+
\delta_{ad}\delta_{bc})+\frac{g_2}{8}(d_{abm}d_{cdm}
+d_{acm}d_{bdm}+d_{adm}d_{bcm}),\\
H_{ab,cd}=&\ \frac{g_1}{4}\delta_{ab}\delta_{cd}+\frac{g_2}{8}(d_{abm}d_{cdm}
+f_{acm}f_{bdm}+f_{adm}f_{bcm}).
\end{split}
\label{Eq:F_H_tensors}
\ee

The sum of the last two terms in the second line of
(\ref{Eq:F_H_tensors}) can be rewritten using the following 
 relation
between the structure constants of the $SU(N)$ group \cite{macfarlane68}
\be
f_{acm}f_{bdm}=\frac{2}{N}\left(\delta_{ab}\delta_{cd}-\delta_{ad}\delta_{bc}
\right)+d_{abm}d_{cdm}-d_{adm}d_{bcm},
\label{Eq:ff_dd_relation}
\ee
resulting in:
\be
H_{ab,cd}=
\frac{1}{4}\left(g_1+\frac{2g_2}{N}\right)\delta_{ab}\delta_{cd}-
\frac{g_2}{4N}\left(\delta_{ac}\delta_{bd}+\delta_{ad}\delta_{bc}\right)
+\frac{3g_2}{8} d_{abm}d_{cdm}-\frac{g_2}{8}\left(
d_{acm}d_{bdm}+d_{adm}d_{bcm}
\right).
\ee

One can express the coupling tensors in terms of the following 
combinations of only six out of the nine independent invariant rank-4 
tensors of the $SU(N)$ group (see eg. \cite{dittner71}):
\begin{alignat}{2}
t^1_{abcd}&=\delta_{ab}\delta_{cd},&\qquad
t^2_{abcd}&=\delta_{ac}\delta_{bd}+\delta_{ad}\delta_{bc},\nonumber\\ 
t^3_{abcd}&=d_{abm}d_{cdm},&\qquad
t^4_{abcd}&=d_{acm}d_{bdm}+d_{adm}d_{bcm}.
\end{alignat}
This set of 4 invariants is closed under multiplication
with the following multiplication table:
\begin{alignat}{2}
t^1*t^1&=(N^2-1)t^1, & \qquad t^1*t^2&=2t^1, \qquad t^1*t^3=0, \qquad
t^1*t^4=2N\left(1-\frac{4}{N^2}\right)t^1, \nonumber\\
t^2*t^2&=2t^2, &\qquad t^2*t^3&=2t^3, \qquad t^2*t^4=2t^4, \quad\  
t^3*t^3=N\left(1-\frac{4}{N^2}\right)t^3,
\label{Eq:SUN_multiplication_table}
\\
\nonumber
t^3*t^4&=N\left(1-\frac{12}{N^2}\right)t^3,& 
t^4*t^4&=2\left(1-\frac{4}{N^2}\right)(2t^1+t^2)+
N\left(1-\frac{16}{N^2}\right)t^3-\frac{8}{N}t^4.
\end{alignat}
In deriving (\ref{Eq:SUN_multiplication_table}) we have used
identities which can be found for example in Appendix A of
\cite{azcarraga98}. 

\paragraph{The special case of $\bm{N=3}$.}
To specify the analysis we first note that in this case $t^4$ is not
an independent invariant because of the relation 
$\displaystyle t^4=(t^1+t^2)/3-t^3$
(see \cite{macfarlane68,dittner71} for its derivation).
Due to this reduction in the set of invariant tensors, the tensorial
coupling structure of the $SU(3)\times SU(3)$ model greatly simplifies:
\begin{alignat}{4}
F^S&=F^P=f(t^1+t^2),\qquad &
f&=\frac{1}{24}\left(6g_1+g_2\right), & & \nonumber\\
H&=h_1t^1+h_2t^2+h_3t^3,\qquad &
h_1&=\frac{1}{8}\left(2g_1+g_2\right), &\qquad h_2&=-\frac{g_2}{8},
\quad h_3 &=\frac{g_2}{2}.
\end{alignat}
These tensors represent a closed set under
multiplication, with a multiplication table which can be read off from
(\ref{Eq:SUN_multiplication_table})
by putting $N=3$. Correspondingly, one can introduce the following
counterterm structures for the Dyson-Schwinger equations:
\bea
&
\delta F^S=\delta f_1^St^1+\delta f_2^St^2+\delta f_3^St^3,\qquad
\delta F^P=\delta f_1^Pt^1+\delta f_2^Pt^2+\delta f_3^Pt^3,\nonumber\\
&
\delta H=\delta h_1t^1+\delta h_2 t^2+\delta h_3t^3.
\label{Eq:SU3_coupling_counterterms}
\eea

In order to find the necessary conditions which determine the
counterterms in the broken symmetry phase we have to investigate the
multiplicity structure in the scalar and pseudoscalar sectors. This is
obtained from the renormalised 2PI-Hartree
gap-equations (\ref{ren-gap}) which in more detail read as
\be
\begin{split}
&(M^2_S)_{ab}=\mu^2\delta_{ab}+4 F^S_{abcd}\left[v_c v_d+
O^S_{c e}O^S_{d e} T_F(\tilde M_{S,e}^2)\right]
+4 H_{ab,cd} O^P_{c e}O^P_{d e} T_F(\tilde M_{P,e}^2),\\
&(M^2_P)_{ab}=\mu^2\delta_{ab}+4 H_{ab,cd}\left[v_c v_d+
O^S_{c e}O^S_{d e} T_F(\tilde M_{S,e}^2)\right]
+4 F_{abcd}^P O^P_{c e}O^P_{d e} T_F(\tilde M_{P,e}^2).
\end{split}
\label{Eq:explicit-gap-equation}
\ee
We assume the presence of scalar condensates belonging to the center
of the group: $(v_3,~v_8)$. At tree (classical) level this background
results in different multiplet structures in the two sectors, but
solving (\ref{Eq:explicit-gap-equation}) iteratively one can show that
the emerging exact mass spectra will have eventually the same
structure in both sectors due to the coupling realised by the tensor
$H$. To be specific, when $v_3\ne0,$ $v_8\ne0$ one has 3 degenerate
doublets in the ``planes'' [1,2], [4,5] and [6,7] and one coupled set
with unequal eigenvalues in the [3,8] ``plane''. These ``planes''
correspond to the pairs of fields [$\pi^+,\pi^-$], [$K^+,K^-$],
[$K^0,\bar K^0$], [$\pi^0,\eta$] in the pseudoscalar sector and to
[$a_0^+,a_0^-$], [$\kappa^+,\kappa^-$], [$\kappa^0,\bar\kappa^0$],
[$a_0^0,\sigma$] in the scalar sector. When there is only one
condensate then the two middle sectors join in a degenerate quadruplet
and the fields of the coupled sector become decoupled mass
eigenstates.  For $v_3=0, v_8\ne 0$ direction ``3'' degenerates with the
first ``plane'' resulting in a $3\oplus 4\oplus 1$ multiplet
structure. For $v_8\ne 0,v_3=0$ directions ``3'' and ``8'' remain as
separate singlets, so that the multiplet structure is $2\oplus 1\oplus
4\oplus 1$.

In view of the known multiplicity structure one finds sufficient
number of independent renormalisation conditions to determine the
remaining six coupling counterterms of
(\ref{Eq:SU3_coupling_counterterms}) by requiring the vanishing of the
(divergent) coefficient of each independent tadpole appearing in the
detailed conditions (\ref{detailed-conditions}).  For the degenerate
sectors, since $O^{S/P}_{m e}=\delta_{m e}$, these conditions result
in the vanishing of the trace of the tensor structure in the curly
bracket of each equation contained in (\ref{detailed-conditions}). For
the coupled sectors one has to impose the vanishing of those
components of the tensor structure in the curly brackets which
correspond to the independent elements of these sectors. Important
simplification occurs when one realises that $\delta F^P=\delta
F^S\equiv \delta F$, since the renormalisation conditions are fully
symmetric under the exchange $P\leftrightarrow S$, in view of
$F^P=F^S$.  Since the number of conditions is sufficiently large all
countercouplings $\delta h_i, \delta f_i, i=1,2,3$ can be determined
after evaluating products of the tensors and counterterm tensors
with help of the multiplication table. Since the number of conditions
is large enough, these equations coincide with the equations one
obtains from the matrix-equations (\ref{4coupling-conditions}),
without investigating the multiplet structure at all:
\be
\begin{split}
\delta f_1=&-8T_d[5f(f+\delta f_1)+f(f+\delta f_2)+4h_1(h_1+\delta
h_1)+h_1(h_2+\delta h_2)+h_2(h_1+\delta h_1)],\\
\delta f_2=&-8T_d[f(f+\delta f_2)+h_2(h_2+\delta h_2)],\\
\delta f_3=&-8T_d[f\delta f_3+h_3(h_2+\delta
h_2)+(h_2+5h_3/6)(h_3+\delta h_3)],\\
\delta h_1=&-8T_d[(4h_1+h_2)(f+\delta f_1)+h_1(f+\delta f_2)+5f(h_1+\delta
h_1)+f(h_2+\delta h_2)],\\
\delta h_2=&-8T_d[h_2(f+\delta f_2)+f(h_2+\delta h_2)],\\
\delta h_3=&-8T_d[h_3(f+\delta f_2)+\delta f_3(h_2+5h_3/6)+
f(h_3+\delta h_3)].
\end{split}
\ee
The coupled set for $\delta f_2$ and $\delta h_2$ is closed in
itself. Its solution is substituted into the other four equations
which then fall into two coupled two-variable equations for $(\delta
f_1, \delta h_1)$ and $(\delta f_3, \delta h_3)$, respectively. 

The analysis above shows that lifting the degeneracy 
of the spectra leads to the
determination of the full set of counter couplings in contrast to the
renormalisation in the symmetric phase of the model. In the latter case the
entire spectra is degenerate and taking the trace results in
$t^3_{abcc}=0$, due to $d_{mcc}=0$. Renormalisability conditions can
be derived from (\ref{Eq:renorm_cond_symphase}) only for two  
linear combinations $\delta f=8(4\delta f_1+\delta f_2)$ and 
$\delta h=8(4\delta h_1+\delta h_2)$. One has
\be
\delta f=-T_d\left[f(f+\delta f)+h(h+\delta h)\right],\qquad
\delta h=-T_d\left[h(f+\delta f)+f(h+\delta h)\right],
\ee 
where we introduced $f=(6g_1+g_2)/3$ and $h=8g_1+3g_2$.

\paragraph{Large N analysis.}
We turn to the analysis of the large $N$ limit. The large $N$ scaling
of the parameters and vacuum condensate strengths is established first
by the requirement that the potential energy density is proportional
to the number of degrees of freedom e.g. $\sim N^2$.

We assume also here that condensates are formed in the center of the
group. In order to find the behaviour of the piece $F_{abcd}v_a v_b
v_c v_d$ of the potential one has to examine the large $N$ behaviour
of the structure constants $d_{abc}$, where at least one of the
indices corresponds to a center element of the Lie algebra. The
generalised Gell-Mann matrices building up the center generators are
labelled as $\lambda_{n^2-1},~n=2,3,..,N$, while the two sequences of
the non-diagonal elements (the analogues of $\sigma^1$ and $\sigma^2$
for $SU(2)$) hold the obvious double index labelling
$\lambda^{\alpha}(i,j),~\alpha=1,2,~i=1,..,N, j=2,..,N, i<j$
\cite{greiner}. One finds the expressions of the structure constants
necessary for our analysis in Table~\ref{d-coeff}. Note
that if one index corresponds to some diagonal $\lambda_{n^2-1}$ then
$d_{ab,n^2-1}$ is nonzero only if $a=b$.

\renewcommand{\arraystretch}{1.35}
\begin{table}
\begin{center}
\begin{tabular}{cccc} \hline 
  $d_{m^2-1,m^2-1,n^2-1}$ & Index relation & $d_{(ij),(ij),n^2-1}$& Index
 relation \\ \hline
  $0$ & $m>n$ & $0$ & $n<i<j$ \\
  $2-m$   & $m=n$ & $\frac{1-n}{2}$ & $n=i<j$ \\
  $1$ & $m<n$ & $\frac{1}{2}$ & $i<n<j$\\
   &  & $\frac{2-n}{2}$ & $i<n=j$ \\
   & & $1$& $i<j<n$ \\ \hline
\end{tabular}
\end{center}
\caption{$SU(N)$ structure constants $d_{ab(n^2-1)}$ in multiples of 
$(2/(n^2-n))^{1/2}$. For the index convention ($a,b=m^2-1, (ij)$) see 
the text. The structure constants are the same for both $\alpha=1,2$.}
\label{d-coeff}
\end{table}
\renewcommand{\arraystretch}{1.0}

The simplest is to analyse the behaviour in presence of a single component
condensate, $v_{n^2-1}$:
\be
V_{tree}=\frac{1}{2}\mu^2 v^2_{n^2-1}+\frac{1}{3}F_{n^2-1,n^2-1,n^2-1,n^2-1}
v^4_{n^2-1}.
\ee
Here
\be
F_{n^2-1,n^2-1,n^2-1,n^2-1}=\frac{3}{4}g_1+\frac{3}{8}g_2\sum_{m=2}^N
d_{n^2-1,n^2-1,m^2-1}d_{n^2-1,n^2-1,m^2-1},
\ee
with
\be
\sum_{m=2}^Nd_{n^2-1,n^2-1,m^2-1}d_{n^2-1,n^2-1,m^2-1}=\frac{2}{n}
\left(1-\frac{n}{N}+\frac{(2-n)^2}{n-1}\right).
\ee
From the quadratic term of the potential one deduces the scaling of
the condensate and using the large $N$ expression of the tensor $F$ from the
quartic term one finds the $N$-scaling of the couplings:
\be
v_{n^2-1}\sim N,~~g_1, g_2\sim \frac{1}{N^2}.
\label{Eq:large_N_scalling}
\ee 
These scaling relations are important when we look for the
non-vanishing tadpole contribution to the gap equations in the large
$N$ limit. They are different from the scaling one finds in the $U(N)\times 
U(N)$ symmetric model for the singlet condensate $v_0$!

Next, one studies the multiplet structure of the model in presence of
the symmetry breaking condensate. We restricted our investigation to
the case of a single component condensate, $v_{n^2-1}$.
There is no mixing between the modes belonging to the different generators.
The tree level masses in the large $N$ limit are given by the following
formulae in the `S'- and `P'-sectors:
\bea
(m^2_S)_{ab}-\mu^2\delta_{ab}&=&
v_{n^2-1}^2\bigg(g_1(\delta_{ab}+2\delta_{n^2-1,a}
\delta_{n^2-1,b})\nonumber\\&&
+\frac{g_2}{2}\sum_c(d_{abc}d_{n^2-1,n^2-1,c}+2d_{a,n^2-1,c}
d_{b,n^2-1,c})\bigg),\nonumber\\
(m^2_P)_{ab}-\mu^2\delta_{ab}&=&v_{n^2-1}^2
\bigg(\Big(g_1+2\frac{g_2}{N}\Big)\delta_{ab}
-2\frac{g_2}{N}\delta_{a,n^2-1}\delta_{b,n^2-1}\nonumber\\
&&+\frac{g_2}{2}\sum_c(3d_{abc}d_{n^2-1,n^2-1,c}-
2d_{a,n^2-1,c}d_{b,n^2-1,c})\bigg).
\eea
The expressions for the masses, which  can be calculated using the structure
constants of Table~\ref{d-coeff}, are given in Table~\ref{N-mass}
together with their multiplicities. The multiplicities take into
account the twofold degeneracy of the ``(ij)'' modes, making in this
way their sum $N^2-1$.

In the generic case, when $n$ and $N-n$ are both ${\cal O}(N)$, one
has three different multiplets of multiplicity ${\cal O}(N^2)$, which
give finite tadpole contribution in the large $N$ limit to the gap
equations. In this case one has enough number of renormalisation
conditions that one can consider the matrix form of the
renormalisation conditions without projection. The other case is when
either $n$, or $N-n$ are ${\cal O}(N^0)$. Then only a single multiplet
of multiplicity ${\cal O}(N^2)$ is formed whose tadpole contributes to
the gap equations of the different fields. In this case one has to
consider appropriate traces in equations (\ref{detailed-conditions})
and (\ref{Eq:explicit-gap-equation}).  One should note that the mass
expressions listed in Table~\ref{N-mass} have different large $N$
values in the two cases: $a\sim {\cal O}(N)$ or $a\sim {\cal O}(N^0)$.

\renewcommand{\arraystretch}{1.35}
\begin{table}
\begin{center}
\begin{tabular}{cccc} \hline 
$(m_{S}^2-\mu^2)_{ab}/v_{n^2-1}^2$ & $(m_{P}^2-\mu^2)_{ab}/v_{n^2-1}^2$ & 
Index $a=b=l^2-1$ & Multiplicity \\ \hline 
$g_1+g_2\big(\frac{3}{l(l-1)}-\frac{1}{N}\big)$ & 
$g_1+g_2\big(\frac{1}{l(l-1)}-\frac{1}{N}\big)$ & $l>n$ & $N-n$ \\
$3\!\Big[g_1+g_2\big(\frac{l^2-3l+3}{l(l-1)}-\frac{1}{N}\big)\Big]$ &
$g_1+g_2\big(\frac{l^2-3l+3}{l(l-1)}-\frac{1}{N}\big)$  & $l=n$ & $1$ \\
$g_1+g_2\big(\frac{3}{n(n-1)}-\frac{1}{N}\big)$ & 
$g_1+g_2\big(\frac{1}{n(n-1)}-\frac{1}{N}\big)$ & $l<n$ & $n-2$\\\hline 
& & Index $a=b=(ij)$&\\ \hline
$g_1-g_2\frac{1}{N}$ & $g_1-g_2\frac{1}{N}$ & $n<i<j$ & $(N-n)(N-n-1)$\\
$g_1+g_2\left(\frac{i-1}{i}-\frac{1}{N}\right)$ & 
$g_1+g_2\left(\frac{i-1}{i}-\frac{1}{N}\right)$ & $n=i<j$ & $2(N-n)$\\
$g_1+g_2\big(\frac{1}{n(n-1)}-\frac{1}{N}\big)$ & 
$g_1+g_2\big(\frac{1}{n(n-1)}-\frac{1}{N}\big)$ & $i<n<j$& 
$2(N-n)(n-1)$\\
$g_1+g_2\big(\frac{j^2-3j+3}{j(j-1)}-\frac{1}{N}\big)$ & 
$g_1+g_2\big(\frac{j^2-j+1}{j(j-1)}-\frac{1}{N}\big)$ &
$i<j=n$ & $2(n-1)$\\
$g_1+g_2\big(\frac{3}{n(n-1)}-\frac{1}{N}\big)$ & 
$g_1+g_2\big(\frac{1}{n(n-1)}-\frac{1}{N}\big)$ & $i<j<n$  & 
$(n-1)(n-2)$\\
\hline
\end{tabular}
\end{center}
\caption{Tree level mass splittings of the $SU(N)\times SU(N)$
  symmetric meson model due to the symmetry breaking condensate
$v_{n^2-1}$ with their respective degeneracies.}
\label{N-mass}
\end{table}
\renewcommand{\arraystretch}{1.0}

Let us start to discuss this last case first. The modes of the 
single contributing multiplet can be completed to the full set 
with no effect on the large $N$ asymptotics. In the resulting 
gap equations we denote its exact mass by $M_{min}$:
\bea
(M_{S/P}^2)_{ab}=(m_{S/P}^2)_{ab}+4(F_{abcc}+H_{ab,cc})T_F(M_{min}).
\eea
In the large $N$ limit the partial trace of the coupling tensors reduces to
\be
F_{abcc}=\frac{g_1}{4}N^2\delta_{ab}+\frac{g_2}{4} d_{acm}d_{bcm},
\qquad
H_{ab,cc}=\frac{g_1}{4}N^2\delta_{ab}-\frac{g_2}{4} d_{acm}d_{bcm}.
\ee
Since $d_{acm}d_{bcm}=(N^2-4)/N \delta_{ab}$, in view of the $N$-scaling
of the couplings (\ref{Eq:large_N_scalling}), in the coupling tensors
above the terms proportional with $g_2$ are subleading relative 
to those with $g_1$. Only the $O(2N^2)$ symmetric coupling survives 
the $N\rightarrow\infty$ limit.

The large $N$ form of the renormalisation conditions is obtained from
(\ref{detailed-conditions}) by taking the trace of the coefficients
multiplying $O_{me}O_{ne} T_F(M_{min})$ in the indices $m,n$. For a
consistent $N$-scaling of the coupling counterterms one has to assume
that the counter couplings $\delta f_i,\delta h_i$ obey the same $\sim
N^{-2}$ scaling like the renormalised couplings. The only entry of
the multiplication table (\ref{Eq:SUN_multiplication_table}) which
produces a factor $\sim N^2$ compensating the quadratic dependence on
the couplings of the terms of (\ref{detailed-conditions}) proportional
to $T_d$ is $t^1*t^1$.  One arrives at the simplified conditions:
\be
\begin{split}
&\delta f_1+4T_dN^2((f_1+\delta f_1)f_1+(h_1+\delta h_1)h_1)=0,\\
&\delta h_1+4T_dN^2((f_1+\delta f_1)h_1+(h_1+\delta h_1)f_1)=0.
\end{split}
\ee
Since at large $N$ $f_1=h_1=g_1/4$, one can consistently choose
$\delta h_1=\delta f_1\equiv\delta g_1$ and renormalise the large $N$ gap
equations of the $SU(N)$ symmetric scalar theory with a single
counterterm (no counterterm has to be introduced to leading order to
$g_2$). This is essentially the large $N$ limit of broken symmetry
phase of the $O(2N^2)$ symmetric model (compare to
(\ref{Eq:ON_largeN_scaling})):
\be
\delta g_1 =-\frac{2N^2 T_d g_1^2}{1+2N^2 T_d g_1}.
\ee

Also in the generic case with three ${\cal O}(N^2)$ size multiplets their gap
equations will depend only on the coupling $g_1$. This is consistent
with the fact that to leading order only terms proportional to
$t^1$ appear in the renormalisation conditions. Therefore only
$g_1$ will renormalise non-trivially, and exactly the same way as
above. One can observe that the masses of the smaller multiplets
depend also on $g_2$ through the tree level contribution. To leading
order $g_2$ is renormalisation invariant, its counterterm will be 
${\cal O}(N^{-3})$.

\section{Conclusions}

In this paper we succeeded to construct explicitly the counterterms of
the 2PI-Hartree approximation to the quantum action of a rather
general class of scalar field theories in a transparent single-step
procedure. The allowed set of counterterms arises by considering all
rank-4 invariant tensors of the actual internal symmetry group of the
theory. Depending on the spectral degeneracy in the broken symmetry
phase, the number of counterterms one actually needs might reduce.
This circumstance is rather peculiar, since in the usual
(non-resummed) perturbative renormalisation the counterterms are
universal, that is independent on any infrared details, especially on
the way the symmetry is broken.  This circumstance is usually
considered to be the consequence of truncating the
2PI effective action. In high enough order of the truncation one
expects that a unique renormalisation scale dependence will be
approached for all 4-point functions defined with functional
derivatives of the effective action with respect to alternative
combinations of fields and propagators.

We have applied this scheme of counterterms to a number of scalar
field theories. For the $O(N)$ symmetric model, to leading order in the
large $N$ approximation, the scheme simplifies to a single quartic
coupling. Its expression coincides with the exact large $N$
asymptotics of the full theory (not truncated at the Hartree level).
The asymptotic expressions for the counterterms of the $SU(N)\times
SU(N)$ theory coincide with the expressions of the $O(2N^2)$ symmetric
model.
 
We plan to investigate the relation of the presented renormalisation
scheme, which requires multiple coupling counterterms with the
symmetric and mass-independent renormalisation which in the
``symmetryzed'' approximation of \cite{ivanov05b} to the $O(N)$ model
was proven to have the counterterm structure expected from the
perturbation theory. Another interesting direction will be to see to
what extent the proposed single-step procedure can be incorporated in
the renormalisation of the momentum-dependent truncations of the 2PI
approximation recently studied in
\cite{heesII02,blaizot04,berges05,toni}.

\section*{Acknowledgements}
Work supported by the Hungarian Scientific Research Fund (OTKA) under
contract no. T046129. Zs. Sz. is supported by OTKA Postdoctoral
Grant no. PD 050015.

\end{document}